# Living Labs – An Ethical Challenge for Researchers and Platform Operators[1]

Philipp Schaer

Institute of Information Science, TH Köln (University of Applied Sciences)

Corresponding author details: Institute of Information Science, Faculty of Information Science and Communication Studies, TH Köln (University of Applied Sciences), Claudiusstr. 1, 50678 Cologne, Germany. E-mail: philipp.schaer@th-koeln.de

## Introduction

The infamous Facebook emotion contagion experiment (Kramer, Guillory, & Hancock, 2014) is one of the most prominent and best-known online experiments based on the concept of what we here call "living labs". In these kinds of experiments, real-world applications such as social web platforms trigger experimental switches inside their system to present experimental changes to their users – most of the time without the users being aware of their role as virtual guinea pigs. In the Facebook example the researches changed the way users' personal timeline was compiled to test the influence on the users' moods and feelings. The reactions to these experiments showed the inherent ethical issues such living labs settings bring up, mainly the study's lack of informed consent procedures, as well as a more general critique of the flaws in the experimental design (Panger, 2016).

While, to the general public, these kinds of experiments were a reason for outrage over the research practices at Facebook, the fact that nearly every commercial digital platform operator is implementing online experiments of many different flavors was not in the center of the still ongoing discussion. Next to social web platform operators like Facebook, search engine operators like Google or Microsoft are also known to massively invest into online experiments and living labs within their platforms and products (Buscher, 2013).

In this chapter, we describe additional use cases to compliment these: The so-called living labs that focus on experimentation with information systems such as search engines and wikis and especially on their real-world usage. The living labs paradigm allows researchers to conduct research in real-world environments or systems. In the field of information science and especially information retrieval – which is the scientific discipline that is concerned with the research of search engines, information systems, and search related algorithms and techniques – it is still common practice to perform *in vitro* or offline evaluations using static test collections. Living labs are widely unknown or unavailable to academic researchers in these fields. A main benefit of living labs is their potential to offer new ways and possibilities to experiment with information systems and especially their users, but on the other hand they introduce a whole set of ethical issues that we would like to address in this chapter.

---

[1] To appear in: Zimmer, M., & Kinder-Kurlanda, K. (Eds.): Internet Research Ethics for the Social Age: New Challenges, Cases, and Contexts. Peter Lang (2017).

Although some questions regarding the ethical challenges that derive from living labs have been discussed before (Sainz, 2012), the topic is relatively new and not much common agreement on the implications and consequences for platform developers exists.

Following an introduction into the field and a broad overview of the possible manifestations of living labs, we present three use cases in this chapter that our research group was involved in over the last years:

- *User motivational studies in a Semantic MediaWiki*. We implemented different user interface variants that we presented to users of a wiki-based research information system. The idea was to trigger different motivation strategies to encourage users to participate more actively in the platform.
- *Ranking studies in an online toy store*. We participated in the 2015 Living Lab for Information Retrieval workshop and retrieval campaign (LL4IR) that allows different research groups to test and compare the effectiveness of their retrieval systems. We computed different ranking results for an online toy store based on historical click data (i.e. usage popularity) and a text-based relevance score (Schaer & Tavakolpoursaleh, 2015).
- *Implementation of a living lab environment into a repository system*. We implemented an application programming interface into an open access repository. The interface allows external collaborators to conduct online retrieval experiments within our own system.

**Enabling In Vivo Experimentation: Living Labs and Innovation Systems**

The core idea of living labs goes back to the 1980s but gained more attention in the years after 2006 when the European Commission started funding the living labs movement (Dutilleul, Birrer, & Mensink, 2010, p. 63). In its core living labs are moving from *in vitro* to *in vivo* research settings by gathering research data from living environments like buildings or public spaces (both online and offline). Researchers are no longer observing artificial laboratory settings but try to observe real-world usage or interactions of people in these research environments that were previously prepared to function as living labs. For example, a store can be wired to record the customers and the employees to learn about their interactions or buying and working patterns.

In market research and innovation management the great potential of these research approaches was picked up quite early and led to including costumers and users early on in the product development process and to more actively involving them in general. As claimed by Chesbrough (2003, p. 41) the traditional model for innovation is becoming obsolete. Innovation is moving from a mainly internal focus, closed off from outside ideas and technologies, to a new paradigm called *open innovation* that is often connected to ideas of living labs and so-called *innovation systems*.

Besides the meaning of living labs as innovation systems, other meanings have evolved over time that are dependent on different foci and use cases. Dutilleul et al. (2010, p. 64) described (among others) four main and distinct meanings of living labs that are present: 1) the already mentioned innovation systems, 2) in vivo monitoring of social systems and the usage of technology within these systems, 3) involving users in the product development process, and 4) organizations developing and maintaining their technological infrastructure and offering services. Living labs are also used in the context of interactive machine learning (Holzinger, 2015), a technique used for content testing at Google, Facebook, and other online platforms. Additional use cases are known to be present in user experience research and ergonomics (Javier & Charles, 2012).

## User Motivational Studies Using Living Labs Approaches

Online communities are highly dependent on their users and their personal involvement in the participation in the community. The activity level is the key concept that makes these communities successful. From the perspective of an online community platform operator we might ask what the central concepts and mechanisms are that make an online community successful. To learn more about the dynamics that drive successful online communities and especially Wiki-based systems we conducted a living labs-based online experiment using the SOFISwiki,[2] a specialized information system that was transformed into a Wiki system.

SOFISwiki is an online platform for a specialized information database that lists social science research projects from Germany, Austria, and Switzerland. The wiki contains more than 50,000 project records from disciplines such as sociology or political sciences. Until the end of 2012 new records for SOFIS were added through an annual survey that was sent out to more than 5,000 social science-related institutions like universities or research and infrastructure institutes. The survey was paper-based and was curated by a team of information professionals and librarians. In late 2012 the new SOFISwiki platform was started, based on a Semantic MediaWiki (SMW) system.

We manipulated the MediaWiki software to include an additional information dashboard right after the users logged into the system. On this dashboard users saw different performance indicators for their own projects and user account (see Figure 1). Each participating user of the system was randomly assigned to one user group. The user groups differed in some aspects of the performance dashboard. During the experiment all activities in the system SOFISwiki of every participating user were logged by the platform to allow us to measure the effects of functionality changes and of different information presented to the users. From the experiment data we were able to positively evaluate the motivational influence of content attribution. So, if users were aware that the good performance of one of their projects in the platform's view statistics was attributed to them (by showing their names and relation), they were more motivated to participate and to contribute more to the platform.

---

[2] http://sofis.gesis.org

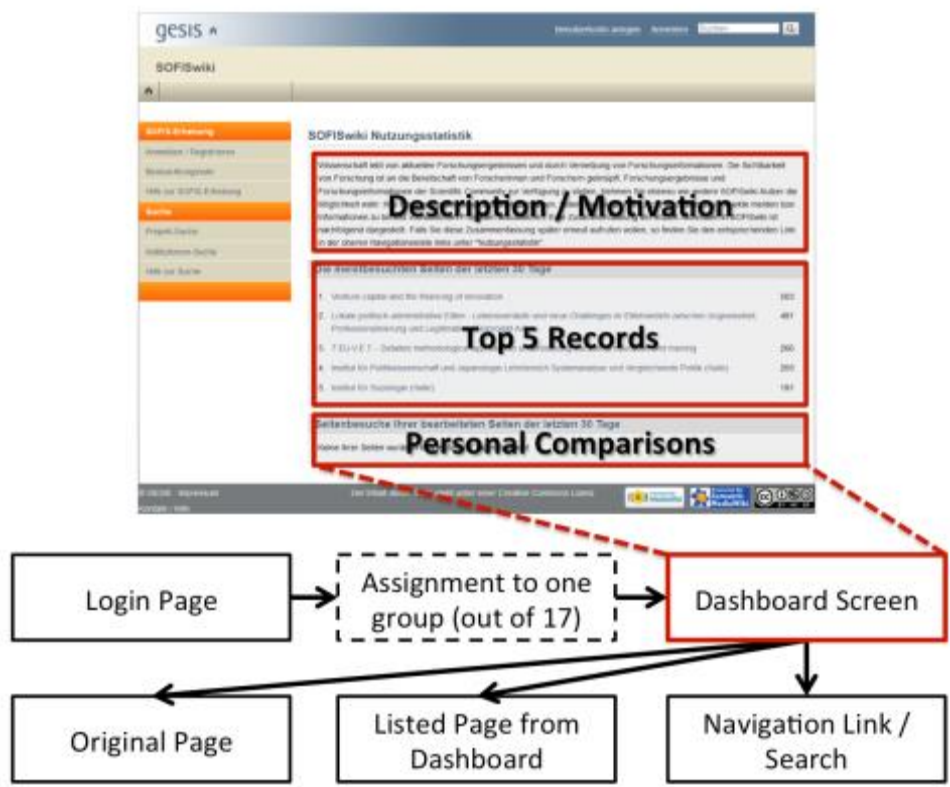

**Figure 1.** Extended login process for the SOFISwiki experiment. After users logged in they were automatically assigned to one group and were redirected to a personalized dashboard where some performance indicators like numbers of view per project were listed

**Ethical Concerns in the Wiki Setup**

Before we started the experiment we released a new version of our terms of use for the SOFISwiki. It included a paragraph on the topic of online experiments and that we would use the interaction data of the platform. This introduces the first ethical issue.

- Although we informed the users about the ongoing (general) changes to the site and the terms of use by showing a notification pop-up, we did not explicitly inform them about every detail of the experiments. But should we not tell the people working on our systems that they are part of an experiment?
- What about the freedom of choice (to not take part in these experiments)? There was no option to opt out of the experiment. Either you were silently accepting the new terms of use and therefore automatically took part in our experiment, or you just had the opportunity to refuse to continue using our system. We did not include a way to simply reject participation in the experiment while still using the platform.
- How much can/should we tell users without spoiling the experiment itself? By telling too much about the motivation behind the experiment itself we would have potentially spoiled the whole experimental setup. Our participants would have been biased because of the background knowledge about our hypotheses and modifications to the system.
- Since Wiki systems tend to be open by design, most of the usage data we showed in the dashboards were publicly available. So the performance indicators are visible as long as you know where to look within the system. We were not able to hide all of this information due to the open nature of the MediaWiki system. Since not every user of the platform uses a pseudonym, one could backlink projects to user accounts and finally to actual persons. A question that might arise here is whether systems that are open by design are a suitable environment for user-centered research, especially when users are not fully aware of the analytical steps that might follow their interactions.

Most of the previously mentioned issues came up late during the experiment or even later during the data analysis phase and we were not able to address these issues at the time when we were running the living lab experiments. We chose to keep a low profile, so we didn't actively announce or comment on the experiment or the ongoing modifications of the platform. This led to a passive behavior or ignorance towards our experiment: The additional steps the users had to perform were silently accepted. Our first level support didn't notice a significantly higher demand for consultation. It seemed as our users just used the platform as usual, regardless of what we were presenting them.

## Living Labs for Information Systems and Retrieval Tasks

The field of information retrieval – summarized as the scientific discipline that is concerned with the research of search engines, information systems, and search related algorithms and techniques – is traditionally heavily dependent on a rigorous empirical evaluation pattern called the Cranfield or TREC paradigm (Sanderson, 2010). In this evaluation technique three essential components are needed: 1) A data set of information objects (most often text documents), 2) a set of topics that describes different information needs, and 3) a set of relevance assessments that judge documents as relevant or non-relevant to the given topics. Obtaining the relevance assessments is a hard and expensive process that relies on human expertise and is an intellectually challenging task. Although some researchers tend to give this task to crowd sourcing platforms like Amazon's Mechanical Turk, most often domain experts are used to judge the search engines results.

To overcome the limitations of the Cranfield paradigm new evaluation techniques were proposed, including online experiments and living labs that allow *in vivo* experiments. Here the meaning of *in vivo* is the possibility for researchers to implement their own algorithms and search methodologies into a live and productive information system. The benefit for the researchers is that they get access to the usage data of real-world users of the information system that are confronted with the results of their search algorithm.

Large search engine companies such as Google or Microsoft use this search engine evaluation technique within their own systems. It has been in use for many years and helps to improve search engines and information systems. The results and the details of these experiments are of course hidden from the public and are only available to the platform developers themselves. Academic researchers are usually not able to participate in these kinds of experiments and to get their hands on the proprietary usage data. By introducing the Living Labs for Information Retrieval initiative (LL4IR) this evaluation paradigm is intended to be open to the academic world. The idea is to find real-world platforms that are willing to implement the LL4IR infrastructure and to open up their platform for the academic community. In this way "living labs are a way to bridge the data divide between academia and industry" (Balog, Elsweiler, Kanoulas, Kelly, & Smucker, 2014). In 2015 the LL4IR initiative received external funding from ELIAS, an ESF Research Networking Programme, and it has established three evaluation campaigns so far.

We were involved within the LL4IR initiative in two different roles: We took part in an evaluation campaign as a research participant, and we implemented the LL4IR API (application programming interface) for our own search engine of the SSOAR system and, therefore, opened up our own system for external research groups (Balog, Schuth, Dekker, Schaer, Chuang, & Tavakolpoursaleh, 2016). The effort to do so was moderate compared to implementing a standalone software solution for different web frameworks and search engines.

**Ethical Challenges for Participants of Living Labs**

In 2015 we took part in the CLEF LL4IR lab (Schaer & Tavakolpoursaleh, 2015). We worked with the online toyshop REGIO JÁTÉK[3] and implemented an alternative product ranking for their search engine that was based on popularity data of the products. The actual ranking mechanism was a combination of a keyword based search and a mixture of word frequency-based relevance scoring and popularity data extracted from clicks provided by REGIO JÁTÉK. So, popular and highly clicked products from the past were more likely to be ranked higher in the result list due to our approach.

The results of our product ranking were presented using an interleaving method called Team Draft Interleaving. This interleaving technique is different to classical online experimental settings like A/B testing as it presents two different rankings at the same time by interleaving the results of two different ranking approaches. Two advantages are apparent: The interleaved presentation of the results lowers the chance of presenting bad results to the user by interleaving experimental (and potentially bad) rankings and rankings produced by the original productive search engine. Another advantage is that by using interleaved comparisons fewer results presentations are needed to evaluate the systems.

Within this evaluation we were confronted with the following ethical questions:

- Is it okay to be biased in your own implementation? By implementing algorithms that are heavily depending on former popularity, how can we suppress a Matthew's effect where only popular content is getting the most attention? This principle is also known as the rich getting richer principle. If one thinks this through, new products or unpopular products never get the chance to be presented to the users in one of the top positions. This is different to letting users explicitly choose between popularity ranking and relevance ranking based on text features.

---

[3]http://www.regiojatek.hu

- We were aware of some issues in our implementation making it a "bad ranking" but nevertheless submitted the results to the LL4IR campaign just to see whether it had any effect or not. Is it ok to present sub-optimal search results to users while the main purpose of an evaluation campaign should be to provide them with the best results possible?

The mentioned issues are softened by using an interleaving method and not A/B testing, but still the potential to do "something evil" is inherent.

**Ethical Challenges for Platform Operators of Living Labs**

After we successfully took part in the evaluation campaign in 2015, we decided to implement the API into our own system, the open access repository SSOAR.[4] We opened the internal search engine for external researchers and took part in the 2016 TREC OpenSearch – Academic Search Edition. Next to CiteSeerX[5] and Microsoft Academic Search[6] we are one of three platforms that provide access to their search engines. By the time of the writing of this article the campaign is still running and no direct results are available. Still, during preparing our system for the TREC OpenSearch, we ran into the following questions:

- Who is responsible for the algorithm and the results? We opened our system and let potential vandals or fanatics present their results. None of the search results provided by the participating researchers are actually controlled by a human assessor.
- What about biased algorithms that present hate-driven, political or religious extremists' content on top of each result list? By opening up the systems one allows others – only on the basis of good will – to decide on the underlying relevance criteria. These might be biased and be based on questionable moral foundations. A ranking algorithm might discriminate in any possible way.

While we believe that the likeliness of such extreme discrimination is very low, we cannot tell if someone will misuse our good intentions. We have no real possibility to check on the validity of the rankings that the external researchers present within our system.

**Conclusions**

We introduced the concept of living labs as a possibility to implement in vivo experiments in real-world information systems like wikis and search engines. We implemented different experimental setups using living labs' principles and encountered a number of ethical questions and issues on the way. We were active in both: conducting research using living lab principles and offering living lab services and interfaces for our own platforms. Both scenarios introduced new insights into the methodology and its ethical drawbacks. Many of these were absolutely new to us, and we had not thought of them before we implemented and executed our experiments. Many of these issues are still open questions and are unresolved. In our experiments we mostly choose to simply ignore a lot of these issues – although we were aware of the possible negative outcomes for individuals or the validity of the experiments. We therefore argue to incorporate ethical concerns and best practices into the research design of living lab-based online experiments since up-to-now these are absolutely not a topic within the community. Nobody seems to care. We will try to bring this discussion to the relevant research community e.g. in the LL4IR initiative.

---

[4] http://www.ssoar.info

[5] http://citeseerx.ist.psu.edu/index

[6] http://academic.research.microsoft.com/

## Acknowledgement

Most of the work presented in this chapter was done during my time with GESIS – Leibniz Institute for the Social Sciences, Cologne, Germany. I would like to thank my former colleagues Narges Tavakolpoursaleh, Simon Bachenberg, Stefan Jakowatz, Annika Marth, and Felix Schwagereit.